\documentclass[prl,aps,showpacs,floatfix,twocolumn,superscriptaddress]{revtex4}
\usepackage{graphicx}
\usepackage[ansinew]{inputenc}
\usepackage{array}
\usepackage{color}
\usepackage{amsmath}
\usepackage{amsxtra}
\usepackage{amstext}
\usepackage{amssymb}
\usepackage{latexsym}
\usepackage{dsfont}
\usepackage{verbatim}

\def\ket#1{\left|#1\right>}
\def\ketbra#1#2{|{#1}\rangle\langle{#2}|}

\begin{document}

\title{All optical control of a single electron spin in diamond}
\author{Y. Chu}
\affiliation{Department of Physics, Harvard University, Cambridge, MA 02138.}

\author{M. Markham}
\affiliation{Element Six, 3901 Burton Drive, Santa Clara, CA 95054}
%

\author{D.J. Twitchen}
\affiliation{Element Six, 3901 Burton Drive, Santa Clara, CA 95054}

\author{M.D. Lukin}
\affiliation{Department of Physics, Harvard University, Cambridge, MA 02138.}

\begin{abstract}
Precise coherent control of the individual electronic spins associated with atom-like impurities in the solid state is  essential for applications in  quantum information processing and quantum metrology. We demonstrate all-optical initialization, fast coherent manipulation, and readout of the electronic spin of the negatively charged nitrogen-vacancy (NV$^-$) center in diamond at T$\sim$7K. We then present the observation of a novel double-dark resonance in the spectroscopy of an individual NV center.  These techniques open the door for new applications ranging from robust manipulation of spin states using geometric quantum gates to quantum sensing and information processing.
\end{abstract}

\maketitle

The negatively charged nitrogen-vacancy (NV$^-$) center in diamond is an atom-like impurity in the solid state that combines a long lived spin-triplet ground state with coherent optical transitions. A number of recent experiments and novel applications have been enabled by the use of a combination of visible frequency lasers to address the electronic states and microwave manipulation to address the spin degree of freedom \cite{DohertyPR2013, PfaffScience2014,  ToganNature2011, RodinArxiv2014, Maurer2012}. While techniques for microwave spin manipulation of NV centers  are well established, a number of new potential applications could be enabled by controlling the ground state spin sublevels using optical Raman transitions as is done with isolated neutral atoms and ions. For example, the use of multiple Raman transitions between four-level tripod systems was proposed for realization of robust geometrical quantum gates \cite{DuanScience2001, MollerPRA2007}.  In addition, optical manipulation offers possibilities for improving NV-based metrology applications, for example by providing access to forbidden transitions between spin sublevels that are more sensitive to magnetic fields \cite{MacQuarrie2013, Fang2013}. Moreover, all-optical manipulation techniques are important for the development of integrated nanophotonic systems for diamond-based scalable quantum optical devices and quantum networks \cite{HausmannNanoLett2013, BurekNL2012, Faraon2011}. In these devices, microwave structures on the diamond substrate are often incompatible with the fabrication process for the nanophotonic devices, while the use of external microwave source defeat the scalability of on-chip photonic devices.  The complex level structure and selection rules of the NV center's optical transitions offer a rich and flexible set of possibilities for coherent all-optical control of all three spin sublevels. Past experiments have demonstrated optical spin manipulation under a large magnetic field or two-photon Rabi oscillations and stimulated Raman adiabatic passage (STIRAP) on microsecond timescales \cite{GolterPRL2014, YalePNAS2013, BassettScience2014}.  

In this Letter, we demonstrate complete all-optical coherent manipulation of the NV spin states.  Importantly, initialization and readout of the spin states are also  performed all-optically, providing a full set of experimental techniques that eliminates the need for microwave addressing. In addition, we report the first observations of nearly degenerate dark states associated with a double-dark resonance in a single quantum emitter \cite{LukinPRA1999, Yelin2003, ChenPRA2001}. Such dark states involving all three sublevels of ground state manifold  open up possibilities for robust control of the entire spin-triplet manifold using geometrical Berry phases \cite{DuanScience2001, MollerPRA2007}.

Our experiments make use of a macroscopic hemisphere of single crystal CVD diamond kept at $\sim$7K in a helium flow cryostat \cite{Sipahigil2012}. The hemisphere acts as a solid immersion lens that increases the efficiency of laser excitation and photon collection. In addition to a 532 nm laser used for spin and charge state initialization, three external cavity diode lasers at 637 nm were used for resonant addressing of various optical transitions. All lasers are pulsed using acousto-optic modulators, with additional electro-optical modulators for generating short pulses with fast rise and fall times on the 637 nm lasers as needed. For initial characterization of the system and during spectroscopy of the tripod system, a permanent magnet outside the cryostat was used to generate a Zeeman splitting of the $\ket{\pm1}$ states. A 15 $\mu$m wire under the bottom face of the sample was used to apply microwave pulses, again only for initial characterization.

\begin{figure}
\centering
\includegraphics[width=0.48\textwidth]{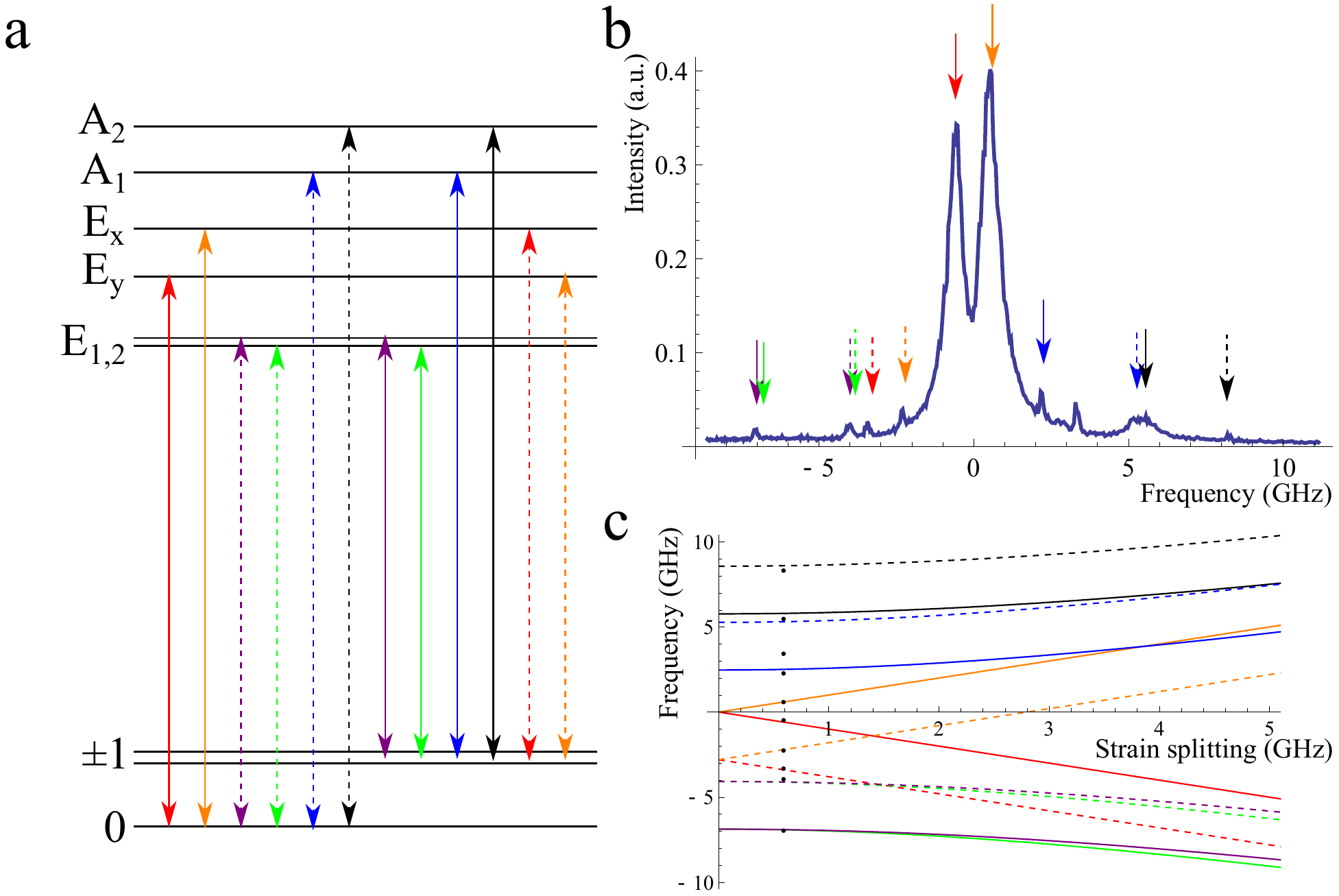}
\caption{Optical transitions of the NV center at 7K. \textbf{a.} All possible transitions between the ground and excited states, with direct transitions indicated with solid lines, and spin non-conserving cross transitions indicated with dashed lines. \textbf{b.} PLE spectrum taken with CW microwave excitation. \textbf{c.} Frequencies of all possible transitions shown in \textbf{a.} as a function of intrinsic crystal strain in units of the splitting between $\ket{E_x}$ and $\ket{E_y}$. The frequencies of the peaks in b. are matched to a particular strain value (black dots). The extra unidentified peak may be due to a two photon transition from the $\ket{0}$ state to the $\ket{E_x}$ state through the absorption of an optical photon and emission of a microwave photon.}
\label{figure1}
\end{figure}

Figure \ref{figure1} presents spectroscopy of the NV center that was used in our experiments. The possible optical transitions between the ground and excited states of the NV center are shown in the level diagram in Figure \ref{figure1}a, including spin-preserving transitions (solid arrows) and non-spin preserving transitions (dashed arrows). All of these transitions can be identified in Figure \ref{figure1}b, where a CW microwave drive was applied between the $\ket{0}$ and $\ket{\pm1}$ ground states and a 637 nm laser was scanned across the zero-phonon transitions while photons were collected on the phonon-sideband (PSB). To further confirm the identification of the observed transitions, we plot in Figure \ref{figure1}c the frequencies of the peaks found in Figure \ref{figure1}b (black dots) on top of the strain dependence of all the optical transitions. The latter was obtained by diagonizing the full Hamiltonian for the optically excited states of the NV center \cite{MazeNJP2011,Rogers:NJP2009}. In the following experiments, we make extensive use of the transitions involving the $A_2$ excited state (black arrows in Figure \ref{figure1}a and b), while spin initialization and readout involve in addition the $\ket{0}\rightarrow\ket{E_x}$ transition. 

To demonstrate all-optical control we first initialize the NV center into an arbitrary ground state spin in zero magnetic field using optical pumping. For example, to prepare the $\ket{-1}$ state, we first apply a laser pulse for 20 $\mu$s on the $\ket{0}\rightarrow\ket{E_x}$ transition. This optically pumps the spin states into $\ket{\pm1}$ through non-spin preserving cross transitions. We then apply a $\sigma_-$ polarized laser to selectively excite the $\ket{+1}\rightarrow\ket{A_2}$ transition for 400 ns. Using this method, we find that a spin polarization of $>$ 80$\%$ in the $\ket{-1}$ state can be achieved. Similarly, the state can be preprepared in the $\ket{+1}$ state by changing the polarization of the laser to $\sigma_-$ in the second step. The effectiveness of this method is limited by off-resonant excitation of the $\ket{\pm1}$ during the first pumping step, decay back into the $\ket{0}$ state during the second pumping step, and imperfect state selection during the second step. In addition, the long optical pumping step on the $\ket{0}\rightarrow\ket{E_x}$ transition can result in ionization of the NV center. We note that the efficiency and fidelity of the optical spin initialization process can be improved by electrically tuning the NV center to give a more optimal combination of cross transition rates \cite{Lucio11}.

In the case of zero Zeeman splitting, it is generally not possible to prepare the NV center in a well-defined superposition of the $\ket{\pm1}$ states using microwave fields since their polarizations are difficult to control. Using optical initialization, however, an arbitrary superposition of the $\ket{\pm1}$ states can be prepared by simply changing the polarization of the laser addressing the $\ket{A_2}$ state. This optical pumping scheme can perform well as long as  Zeeman splitting is smaller than the lifetime of the $\ket{\pm1}\rightarrow\ket{A_2}$ transition. At even higher fields, preparation in either the individual $\ket{+1}$ or $\ket{-1}$ states would be unaffected, while preparation in an arbitrary superposition is possible with two phase-locked lasers at different frequencies. 

The second element in all optical control is the ability to read out any arbitrary spin state. The population of the $\ket{0}$ state can be read out using off-resonant excitation, while resonant single-shot readout of the $\ket{0}$ state has also been demonstrated \cite{Lucio11}. To measure the population in the $\ket{\pm1}$ states, one can first  transfer it to the $\ket{0}$ state using e.g. microwave manipulation. A similar technique could be adopted for all-optical readout by coherently transferring $\ket{\pm1}$ to $\ket{0}$ using Raman transitions (see discussion below).  Here, however, we use a simpler method of directly measuring the $\ket{\pm1}$ population. 
\begin{figure}
\centering
\includegraphics[width=0.48\textwidth]{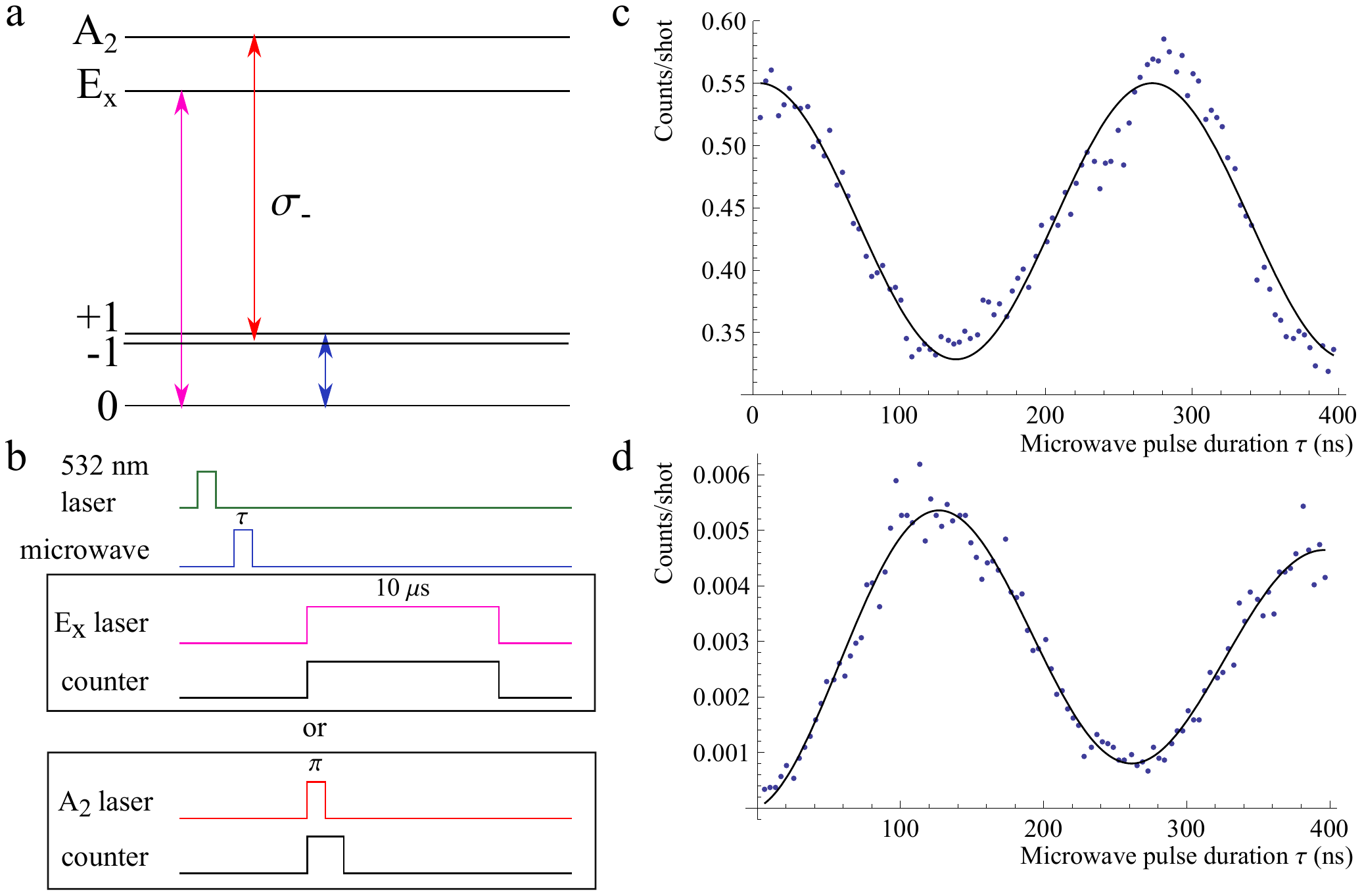}
\caption{Readout of $\ket{-1}$ spin state popluation. \textbf{a.} Level structure showing laser excitation and microwave field. A small external field of B$\sim$10 G was applied. However, the Zeeman splitting is not resolvable with the $\ket{+1}\rightarrow\ket{A_2}$, so state selectivity is ensured by making the laser $\sigma_-$ polarized. \textbf{b.} Pulse sequence showing two alternative readout schemes. \textbf{c.} Ground state Rabi oscillations detected using conventional spin readout with $\ket{0}\rightarrow\ket{E_x}$ transition. \textbf{d.} Ground state Rabi oscillations detected using $\ket{+1}\rightarrow\ket{A_2}$ transition.}
\label{figure2}
\end{figure}
Since, unlike the $\ket{m_s=0}$ states, the transitions involving the $\ket{m_s=\pm1}$ states are not cycling, a long readout pulse will quickly optically pump the population into the orthogonal state in the $\ket{\pm1}$ manifold. Therefore, we instead read out the spin state by applying an optical $\pi$ pulse to the $\ket{A_2}$ state and collecting the emitted photons. We can choose a particular superposition of $\ket{+1}$ and $\ket{-1}$ by choosing the polarization of the readout laser. To separately characterize the effectiveness of this method, we first apply a magnetic field and use this technique to detect microwave-driven Rabi oscillations between $\ket{+1}$ and $\ket{0}$. Figure \ref{figure2} shows a comparison of two methods for reading out the $\ket{+1}$ and $\ket{0}$ populations, along with the associated level scheme and pulse sequences. As expected, the oscillations are out of phase from each other. 
We see that, in comparison with conventional resonant spin readout, the $\ket{+1}\rightarrow\ket{A_2}$ readout results in good contrast, but much fewer counts. 

\begin{figure}
\centering
\includegraphics[width=0.48\textwidth]{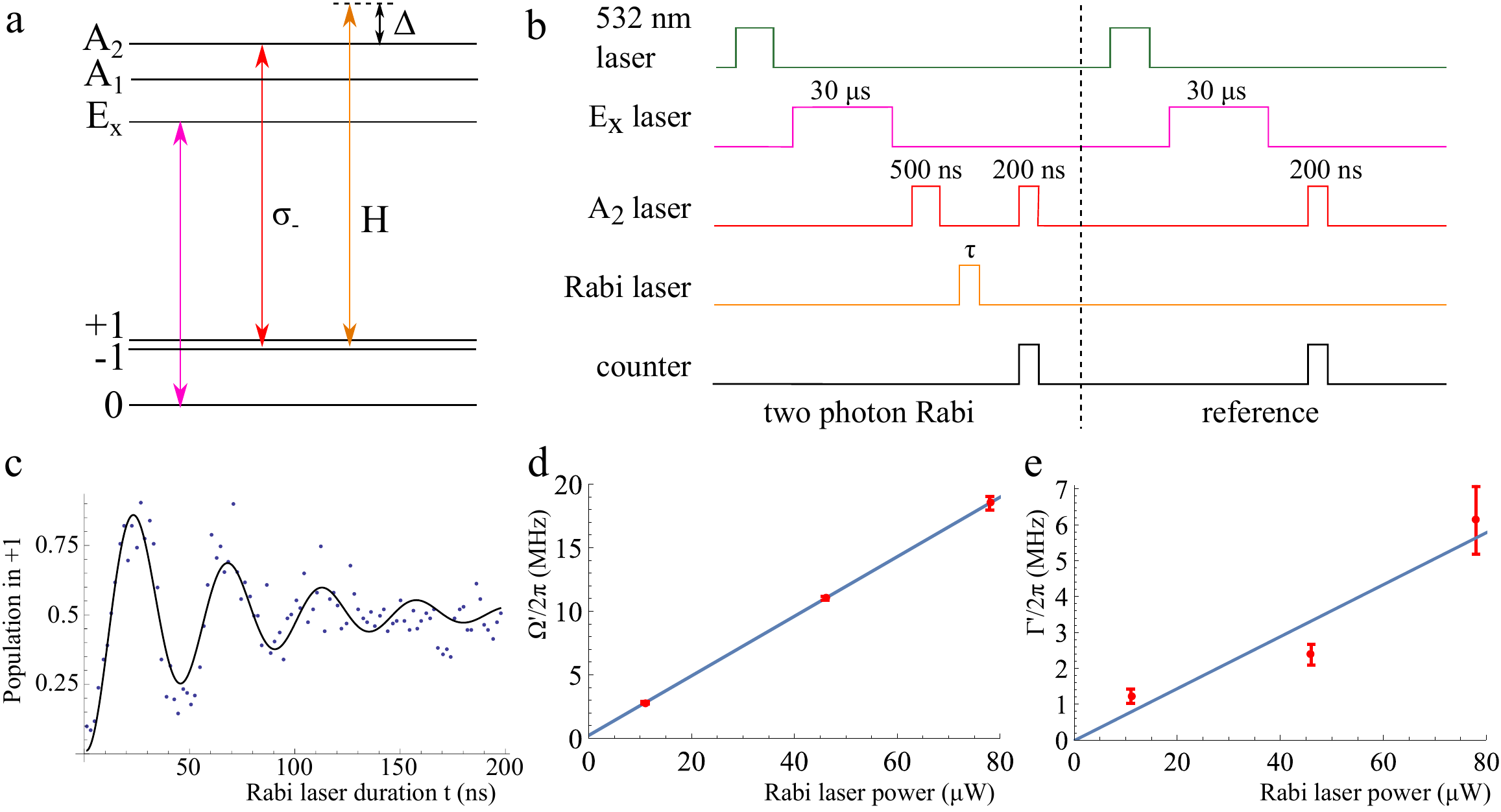}
\caption{Two-photon Rabi oscillations between the $\ket{\pm1}$ ground states. \textbf{a.} Level structure showing the transitions involved. One laser is used for optical pumping from $\ket{0}$ state (magenta). Another $\sigma_-$ polarized laser is used for optical pumping and spin readout from $\ket{+1}$ state (red). A third, linearly polarized laser is used for driving the two-photon transition (orange). \textbf{b.} Pulse sequence showing spin preparation, manipulation, and readout steps. As a reference, the $\ket{+1}$ state was read out after applying only the $\ket{0}\rightarrow\ket{E_x}$ optical pumping step, which prepares the spin state in an equal superposition of $\ket{+1}$ and $\ket{-1}$. \textbf{c.} Two-photon Rabi oscillations with 46 $\mu$W of laser power.  The counts obtained during the readout after two-photon Rabi are normalized to the counts obtained during readout after the reference sequence. \textbf{d.} Two-photon Rabi frequency as a function of laser power, with linear fit (blue line). \textbf{e.} Decay rate of two-photon Rabi oscillations as a function of laser power, with linear fit (blue line).}
\label{figure3}
\end{figure}

We now combine the spin polarization and readout techniques above to demonstrate two-photon Rabi oscillations, see  Figure \ref{figure3}. These experiments are conducted at zero magnetic field, such that selective microwave excitation is not possible. First, the NV center is optically pumped into the $\ket{-1}$ state as described above. Using an EOM, we then apply a linearly polarized laser pulse of varying duration that is detuned $2\pi \times 2.24$ GHz from the $\ket{\pm1}\rightarrow\ket{A_2}$ transition. The population in the $\ket{+1}$ state is then measured  with the same laser as the one used for $\ket{+1}$ to $\ket{-1}$ optical pumping. 
As can be seen in Figure \ref{figure3}c, we observe oscillations of the $\ket{+1}$ state population that eventually decays to a steady state value of 1 relative to the reference, which corresponds to 50$\%$ population. Here, the two-photon Rabi frequency is large enough to drive all hyperfine levels of the ground state, which results in the much higher oscillation contrast compared to previous work \cite{GolterPRL2014}. 

In the case where the two branches of the $\ket{\pm1}\rightarrow\ket{A_2}$ transition are addressed by laser fields $\Omega_+$ and $\Omega_-$ with large one-photon detuning $\Delta$ but zero two-photon detuning, the system behaves as if the $\ket{\pm1}$ states are coupled by an effective two-photon Rabi frequency $\Omega'=\Omega_+^*\Omega_-/|\Delta|$. In our experiments,  a single linearly polarized laser provides both circularly polarized driving fields. Two-photon Rabi oscillations occur in the limit where the detuning $\Delta$ is large enough so that 
$\frac{\Omega_+^2}{\Delta^2}\gamma,\frac{\Omega_-^2}{\Delta^2}\gamma\ll\Omega'$,
where $\gamma$ is the lifetime of the excited state. Additionally, one has to account for the presence of the $\ket{\pm1}\rightarrow\ket{A_1}$ state with an additional detuning of $2\pi\times 3.2$ GHz. Since the $\ket{A_1}$ state has the opposite relative phase as the the $\ket{A_2}$ state between the $\ket{\pm1}$ spin components, we find that the two-photon Rabi frequency is given by
\begin{equation}
\Omega'=\Omega_+^*\Omega_-\left(\frac{1}{\Delta_2}-\frac{1}{\Delta_1}\right)
\end{equation}
where $\Delta_{1,2}$ are the detunings from the $\ket{A_{1,2}}$ states, respectively. In Figure \ref{figure3}d, we perform a linear fit to the power dependence of $\Omega'$. This allows us to extract the corresponding values of $\Omega_{\pm}$, which are in good agreement with independently measured Rabi frequency of the $\ket{\pm1}\rightarrow\ket{A_2}$ transition.

\begin{figure}
\centering
\includegraphics[width=0.48\textwidth]{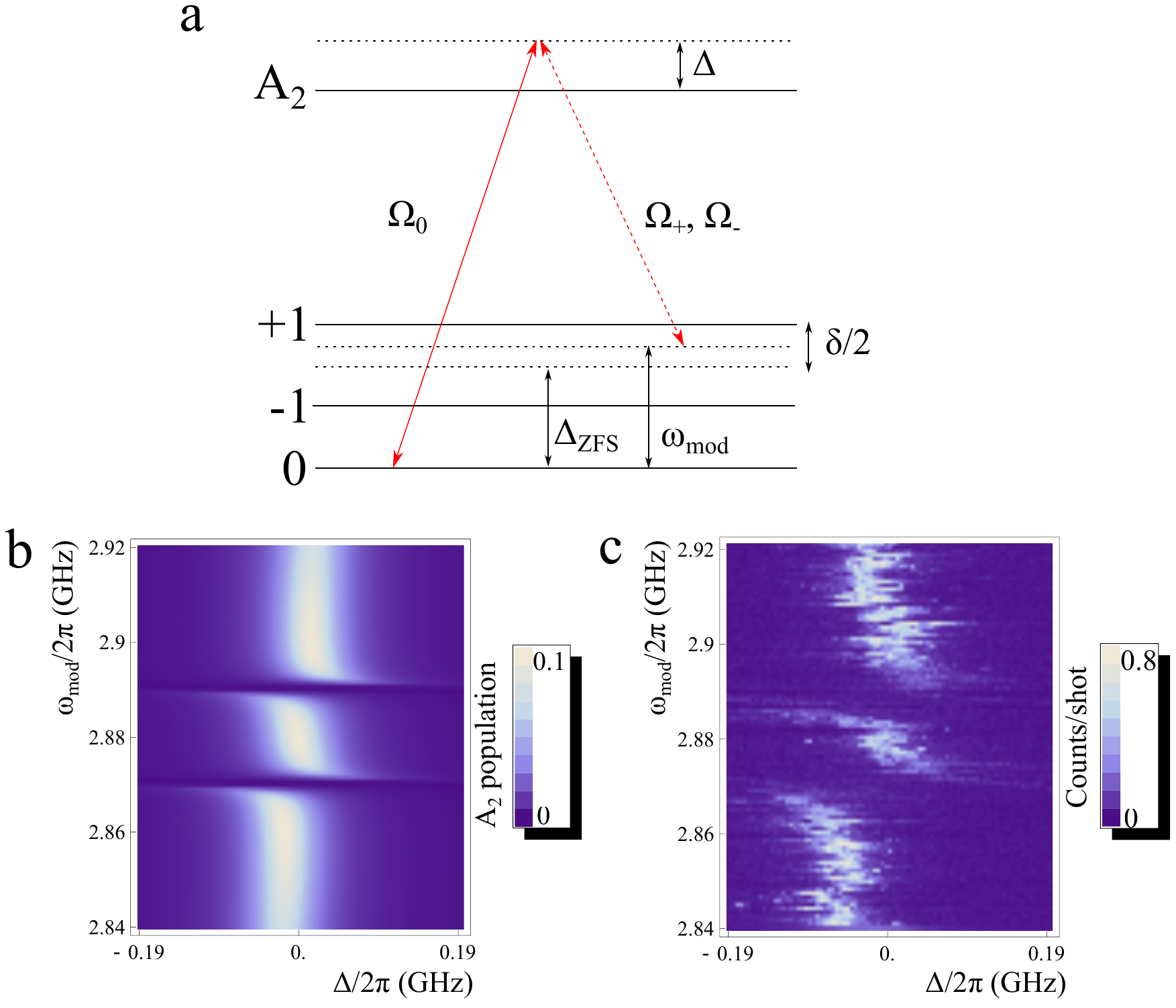}
\caption{Observations of double dark resonances in individual NV centers \textbf{a.} Tripod level structure. \textbf{b.} Theoretical excited state population as a function of one-photon detuning and modulation frequency. \textbf{c.} Experimental data showing PSB fluorescence during laser excitation as a function of one-photon detuning and modulation frequency.}
\label{figure4}
\end{figure}

The decay of the two-photon Rabi oscillations is due several effects. First, for small two-photon Rabi frequencies, the hyperfine splitting due to the $^{14}$N and $^{13}$C spin bath results in decoherence of the Rabi oscillations.  Second, off-resonant excitation leads to spontaneous emission from the excited state, leading to decay of the Rabi oscillations at a rate $\gamma\left(\Omega_+\Omega_-/\Delta^2\right)$. Third, a combination of spectral diffusion of the $\ket{A_2}$ state and frequency fluctuations of the laser gives rise to an uncertainty in the detuning $\delta_{\Delta}$. This last effect is dominant in our case. The total decay rate of the two-photon Rabi oscillations is then given by
\begin{equation}
\Gamma=\left|\Omega_+\right|\left|\Omega_-\right|\left(\frac{1}{\Delta_2^2}-\frac{1}{\Delta_1^2}\right)\delta_{\Delta}.
\end{equation}
Using this expression and the measured values of $\Omega'$ in Figure \ref{figure3}d, we extract a value of $\delta_{\Delta}=2\pi\times490$ MHz from a linear fit to the decay rates (Figure \ref{figure3}e), consistent with independent optical spectroscopy measurements.


We next explore  coherent transfer of population between the $\ket{0}$ and $\ket{\pm1}$ states. This requires the coherent control of non-spin-preserving cross transitions between the ground and excited states \cite{SantoriPRL2006, YalePNAS2013}. 
We present spectroscopy and characterization of the tripod system formed between the $\ket{A_2}$ excited state and all three ground state levels in the same low-strain NV center as above. Remarkably, such a system allows for the possibility of multiple dark resonances associated with transitions linking different all spin sublevels in the ground state manifold. 

To demonstrate optically induced coherence between all three ground state spin levels, we scan a single linearly polarized laser across the $\ket{0}\rightarrow\ket{A_2}$ transition and modulate the same laser using an EOM to create sidebands for addressing the $\ket{\pm1}\rightarrow\ket{A_2}$ transition. We use a 50 $\mu$s excitation pulse and collect PSB photons during that interval. In addition, during each successive scan, we change the modulation frequency such that the sideband sweeps through the $\ket{\pm1}\rightarrow\ket{A_2}$ transition. An external magnetic field was applied to split the $\ket{\pm1}$ states by $2\pi\times 18$ MHz. 
As can be seen in Figure \ref{figure4}c, when the sideband is far detuned from the $\ket{\pm1}\rightarrow\ket{A_2}$ transitions, we simply obtain a resonance corresponding to the $\ket{0}\rightarrow\ket{A_2}$ cross transition. However, when the modulation frequency is such that the $\Lambda$ systems involving the $\ket{0}$ and $\ket{+1}$ or $\ket{0}$ and $\ket{-1}$ states are in two-photon resonance, we observe a decrease in fluorescence corresponding to two nearly degenerate dark states.  

These observations can be understood by using simple tripod model, described by the following Hamiltonian for the four level system: 
\begin{eqnarray}
H&=&-\Delta\ketbra{0}{0}-(\Delta'+\frac{\delta}{2})\ketbra{+1}{+1}-(\Delta'-\frac{\delta}{2})\ketbra{-1}{-1}\nonumber\\
&&-(\Omega_0\ketbra{A_2}{0}+\Omega_+\ketbra{A_2}{+}+\Omega_-\ketbra{A_2}{-}+h.c.).
\end{eqnarray}
Here, as illustrated in Figure \ref{figure4}a, $\Delta$ is the one photon detuning of the carrier laser frequency, $\delta$ is the Zeeman splitting between the $\ket{\pm1}$ states, and $\Delta'=\Delta+\Delta_{ZFS}-\omega_{mod}$, where $\Delta_{ZFS}=2\pi\times2.88$ GHz is the ground state zero field splitting and $\omega_{mod}$ is the laser modulation frequency. $\Omega_0$ is the the Rabi frequency of the laser addressing the $\ket{0}\rightarrow\ket{A_2}$ transition. Importantly, the two dark states associated with this model are 
\begin{equation}
\ket{D_{\pm}} = \frac{\Omega_0\ket{\pm1}-\Omega_\pm\ket{0}}{\sqrt{\left|\Omega_0\right|^2+\left|\Omega_\pm\right|^2}}
\end{equation}
when $\Delta_{ZFS}-\omega_{mod}=\mp\delta/2$.
The systems stops absorbing light whenever either of these two dark resonance conditions is satisfied, regardless of the Zeeman splitting $\delta$. This is the essence of double-dark resonances \cite{LukinPRA1999}.  We can model the system including the effects of excited state decay using a full master equation approach. The results are shown in Figure \ref{figure4}b, and agree qualitatively with the experimental data. Remarkably, two nearly degenerate dark lines separated by the Zeeman splitting are observed, corresponding to double-dark resonances involving the states $\ket{D_+}$ and $\ket{D_-}$ . 
Since the optical transitions between $\ket{A_2}$ and $\ket{\pm1}$ have orthogonal circular polarizations, it is possible to achieve selective and independent coherent operations between $\ket{0}$ and either $\ket{+1}$ or $\ket{-1}$ even at zero magnetic field by choosing the appropriate laser detunings and polarizations. 

The present observations open up new possibilities for the development of a full set of techniques for coherent control of the entire ground state manifold of the NV center using all-optical methods. We have shown that, using enhanced excitation and photon collection associated with SIL, optical manipulation of the spin states can be achieved with comparable efficiency to microwave manipulation. Moreover, the four-level system investigated here can also be used for transferring population between the $\ket{0}$ and $\ket{\pm1}$ states using, for example, two-photon Rabi or STIRAP techniques. In particular, such a configuration enables a universal set of geometric gates on the $\ket{\pm1}$ qubit states by using the $\ket{0}$ state as an ancilla \cite{DuanScience2001, MollerPRA2007}.  These techniques can be immediately applicable to experiments involving 
 NV centers in photonic structures such as optical cavities and waveguides \cite{Faraon2011, HausmannNanoLett2013, BurkardArxiv2014}. With the development of diamond-based photonic devices that further enhance NV-light interactions, the techniques demonstrated can become important elements of integrated quantum network nodes based on NV centers in nanocavities.

%

The authors would like to thank Emre Togan, Alp Sipahigil, Brendan Shields, and Alexander Zibrov for fruitful discussions and technical help. This work was supported by the NSF, CUA, DARPA, MURI, Packard Foundation, ERC Advanced Investigator Grant, and Element Six. Y.C acknowledges financial support from the NSF GRFP.

\bibliography{allOpticalSpin09_21_2014}

\end{document}